\documentclass[12pt,preprint]{aastex}
\usepackage{epsfig}
\slugcomment{submitted to \apj Letters}

\shorttitle{Progenitors for SNe Ia  \& Binary evolution}
\shortauthors{Chen and Han }

\begin{document}
\title{Tidally enhanced stellar wind: a way to make the symbiotic channel to Type Ia supernova viable}


\author{X. Chen, Z. Han and C. A. Tout}
\affil{National Astronomical Observatories / Yunnan
Observatory, CAS, Kunming, 650011, China\\
       Key Laboratory for the Structure and Evolution of Celestial
Objects, CAS\\
Institute of Astronomy, Madingley Road, Cambridge CB3 0HA, England} \email{xuefeichen717@hotmail.com}

\begin{abstract}
In the symbiotic (or WD+RG) channel of the single degenerate
scenario for type Ia supernovae (SNe~Ia) the explosions occur
a relatively long time after star formation.  The birthrate from this
channel would be too low to account for all observed SNe~Ia were
it not for some mechanism to enhance the rate of accretion on to
the white dwarf.  A tidally enhanced stellar
wind, of the type which has been postulated to explain many
phenomena related to giant star evolution in binary systems, can
do this.  Compared to mass-stripping this model
extends the space of SNe~Ia progenitors to longer orbital periods
and hence increases the birthrate to about $0.0069\,{\rm yr^{-1}}$
for the symbiotic channel.  Two symbiotic stars, T~CrB and RS~Oph,
considered to be the most likely progenitors of SNe~Ia
through the symbiotic channel, are well inside the
period--companion mass space predicted by our models.
\end{abstract}

\keywords{binaries: close --- stars: evolution --- supernovae:
general}

\section{Introduction}
Type~Ia supernovae (SNe~Ia) are used as distance indicators for cosmology.  From SNe~Ia we
deduce that the expansion of the Universe is accelerating
\citep{riess98,per99}.  They are also crucial for the study of
Galactic chemical evolution because they expel
iron.  However, the exact nature of their progenitors remains unclear and it is still uncertain whether a
double-degenerate (DD) or single-degenerate (SD) scenario dominates.
In the DD scenario two carbon-oxygen (CO) white dwarfs (WDs) can
produce a SN~Ia while merging if their total mass is larger than
$1.38\,M_\odot$.  In the SD scenario, a CO WD explodes as a SN~Ia if
its mass reaches $1.38\,M_\odot$ while accreting from a non-degenerate
companion such as a main-sequence (MS) star or
slightly evolved subgiant in the WD+MS channel, a giant star in the
WD+RG channel or the symbiotic channel or a helium star
\citep{heu92,han04}.

A progenitor model can be tested by comparing the predicted
distribution of the time between star formation and SN~Ia
explosion with that observed.  This ranges from less than $10^8\,$yr
to more than $3\,$Gyr.  A significant number of SNe~Ia have taken
the longest times \citep{bot08,tot08,sch09}.  Though the DD scenario
can reproduce such delays the SD scenario cannot, except in the study
by \citet{hac08} who assumed that the wind from the WD strips the
outer layers of the red giant (RG) at a high rate (their
mass-stripping effect).  They showed that symbiotic stars are
likely progenitors of SNe~Ia with long delays.  \citet{gilf10}
have claimed that supersoft X-ray (SSX) fluxes in early type galaxies
are much smaller than expected if the SD scenario were to dominate
SNe~Ia production.  However \citet{hac10} argue that this SSX flux is
actually rather strong evidence in favour of the SD scenario in early
type galaxies and that the progenitors are symbiotic stars.

There are at least four recurrent symbiotic novae, such as RS~Oph
or T~CrB, that have a very massive WDs.  They are thought to have
red giant companions of mass $M<1\,M_\odot$ \citep{anu99}.  It is
difficult to reproduce these systems in standard binary evolution
because dynamical instability in mass transfer sets in when the
giant fills its Roche lobe.  \citet{hac99a} systematically studied
binary evolution and established WD+MS and WD+RG paths to SNe~Ia
by introducing three new physical effects, (i) the wind from the
RG acts like a common envelope to reduce the separation for very
wide binaries with separations up to about $30{,}000\,R_\odot$,
(ii) the WD loses much of the transferred mass in a massive
optically thick wind and (iii) their mass-stripping effect. They
produced SNe~Ia at a rate comparable with that observed. {\bf
Since then the only work to produce sufficiently high SN Ia rates
with long delay times was \citet{hac08}, who implemented all the
three effects. Other works adopted the above effects except
mass-stripping in their population syntheses and found smaller SN
Ia rates.}

The concept of a compact object accreting from the stellar wind
of a companion was used by \citet{davidson1973} to
explain X-ray pulses in massive X-ray binaries.  It is now
generally included in population synthesis calculations such as
those by \citet{hurley2002}, though \citet{hac99a} did not
include any wind accretion in their model.  The
surfaces of red giant stars are not tightly bound and so it is
relatively easy to drive a cool wind.  Such winds might be enhanced
by tidal or other interaction with a companion \citep{tout88a} and
so could remove significant mass and angular momentum from the
giant's envelope before Roche lobe overflow (RLOF) begins.
\citet{tout88a} introduced this concept of companion reinforced
attrition (CRAP) to explain mass inversion in RS CVn binaries and
it has since been widely used to explain phenomena related to
giant star evolution in binary systems, such as double degenerate
objects, post-AGB stars, barium stars, cataclysmic variable stars and
bipolar planetary nebulae \citep[see for
example][]{han98,winck03,bon08}.  As potential progenitors of
SNe~Ia symbiotic stars \citep{kenyon1984} gain two advantages
from CRAP.  First the WD may grow
in mass substantially by accretion from stellar wind before RLOF
and secondly mass transfer may be stabilized because the mass
ratio ($M_{\rm giant}/M_{\rm WD}$) can be much reduced at the
onset of RLOF.  So the distribution of masses and periods from
which SNe~Ia arise through the symbiotic channel is enlarged when
CRAP is taken into account.  Here we study this
for SNe~Ia with CRAP instead of \citet{hac99a}'s mass stripping
effect.  Although both processes may occur in real systems we
include only CRAP so as to isolate its effects.

\section{Binary Evolution Calculations and Results}

We evolve about 2,000 population~I (with metallicity $Z = 0.02$) close
WD+MS binary stars with Eggleton's stellar evolution code
\citep{egg71,pol98}.  Their CO WD masses $M_{\rm WD}$ are 0.6, 0.65,
0.8, 1.0 and $1.1\,M_\odot$, the initial orbital periods are between
10 and 10,000\,d spaced at intervals of $\Delta\log_{10}(P^{\rm
  i}/{\rm d})=0.1$ and the companion, secondary masses $M_2$ in the
range from 0.8 to $7.0\,M_\odot$.  The tidally enhanced mass-loss rate
from the secondary $\dot M_{\rm 2w}$ is modelled by
\citet{reimers1975}' formula with the extra tidal term included by
\cite{tout88b} so that
\begin{equation}
\dot M_{\rm 2w}=-4 \times 10^{-13} \frac{\eta
(L/L_\odot)(R/R_\odot)}{(M_{\rm 2}/M_\odot)}\left[1+B_{\rm
W}\,{\rm min}\left(\frac{1}{2}, \frac{R}{R_{\rm
L}}\right)^6\right]\,M_\odot\,{\rm yr}^{-1},
\end{equation}
where $R$ and $L$ are the radius and luminosity of the giant
secondary, $R_{\rm L}$ is its Roche lobe radius, $\eta$ is Reimers'
wind coefficient.  We set $\eta=0.25$.  The wind enhancement parameter
$B_{\rm w}$ we set to 10,000 as required to fit Z~Her \citep{tout88a}.
Thus the mass-loss rate $|\dot M_{\rm 2w}|$ could be 150 times as
larger than Reimers' rate when the star more than half fills its Roche
lobe.

Some of the mass lost in the stellar wind of the giant
may be accreted by the WD at a rate \citep{boff88} so that
\begin{equation}
\label{bhacc}
\dot M_{\rm 2a}=-\frac{1}{\sqrt{1-e^{2}}}\left(\frac{GM_{\rm
WD}}{v_{\rm w}^{2}}\right)^{2} \frac{\alpha_{\rm acc}\dot M_{\rm
2w}}{2a^{2}(1+v_{\rm orb}^{2}/v_{\rm w}^{2})^{3/2}},
\end{equation}
where $v_{\rm orb}=\sqrt{G(M_{\rm 2}+M_{\rm WD})/a}$ is the orbital
velocity, G is Newton's gravitational constant, $a$ is the semi-major
axis of the orbit and $e$ is its eccentricity.  We take $e = 0$
because we expect orbits to have circularized by this time.  The
coefficient $\alpha_{\rm acc}$ is an accretion efficiency and we set
$\alpha_{\rm acc}=1.5$.  We fix the wind velocity $v_{\rm w}$ to
$5\,{\rm km\,s^{-1}}$ and discuss this later.  If $v_{\rm w}$ or $a$
is small, the right hand side of equation~(\ref{bhacc}) becomes large
and so we add the limit $\dot {M}_{\rm 2a} \le -\dot M_{\rm 2w}$.
That part of the wind which is not accreted carries off the specific
angular momentum of the donor star.

Before RLOF begins stellar wind is the
only way to transfer material to the WD so the mass transfer
rate, $\dot M_{\rm t}=\dot M_{\rm 2a}$.  During RLOF the
mass is transferred in both a stream and the wind so that $\dot M_{\rm t} =\dot
M_{\rm 2a}+|\dot M_{\rm 2RLOF}|$, where $\dot M_{\rm 2RLOF}$ is
mass-transfer rate by RLOF.

We continue the practice of limiting the rate at which the WD can
accrete \citep{han04,meng09,wang09} by the prescription of
\citet{hac99b}.  This accounts for the limited rate at which hydrogen
can burn and for mass loss in both hydrogen and helium novae.  Thus
\begin{equation}
\dot M_{\rm WD}=\eta_{\rm H}\eta_{\rm He}\dot M_{\rm t}.
\end{equation}
Hydrogen accretion is controlled by
\begin{equation}
\label{mdotcrit1}
 \eta_{\rm H} = \cases{ \dot M_{\rm cr}\over
|\dot M_{\rm t}|, & $|\dot{M_{\rm t}}|>\dot{M}_{\rm cr}$,\cr 1, &
$\dot{M}_{\rm cr}\ge |\dot{M_{\rm t}}|\ge \dot{M}_{\rm low}$,\cr
0, & $|\dot{M_{\rm t}}|< \dot{M}_{\rm low}$, }
\end{equation}
where $\dot M_{\rm low}$, equal to $\frac{1}{8}\dot{M}_{\rm cr}$,
is the accretion rate below which hydrogen novae expel most of the
material
and $\dot M_{\rm cr}$
is the critical accretion rate above which hydrogen cannot burn as
it is accreted but is instead expelled in an optically thick wind.
Helium accretion is then further controlled by
\begin{equation}
\label{mdotcrit2}
\eta_{\rm He} = \cases{ -0.175(\log_{10}(\dot
M_{\rm He}/M_\odot\,{\rm yr^{-1}}) + 5.35)^2 + 1.05, & $-7.3 <
\log_{10}(\dot M_{\rm He}/M_\odot\,{\rm yr^{-1}}) < -5.9$,\cr 1, &
$-5.9 \le \log_{10}(\dot M_{\rm He}/M_\odot\,{\rm yr^{-1}}) < -5,$
}
\end{equation}
where $\dot M_{\rm He} = \eta_{\rm H}|\dot M_{\rm t}|$.  That part of
the mass transferred but ultimately not accreted carries off the
specific orbital angular momentum of the white dwarf.

\begin{figure}
\includegraphics[width=8.5cm,angle=270]{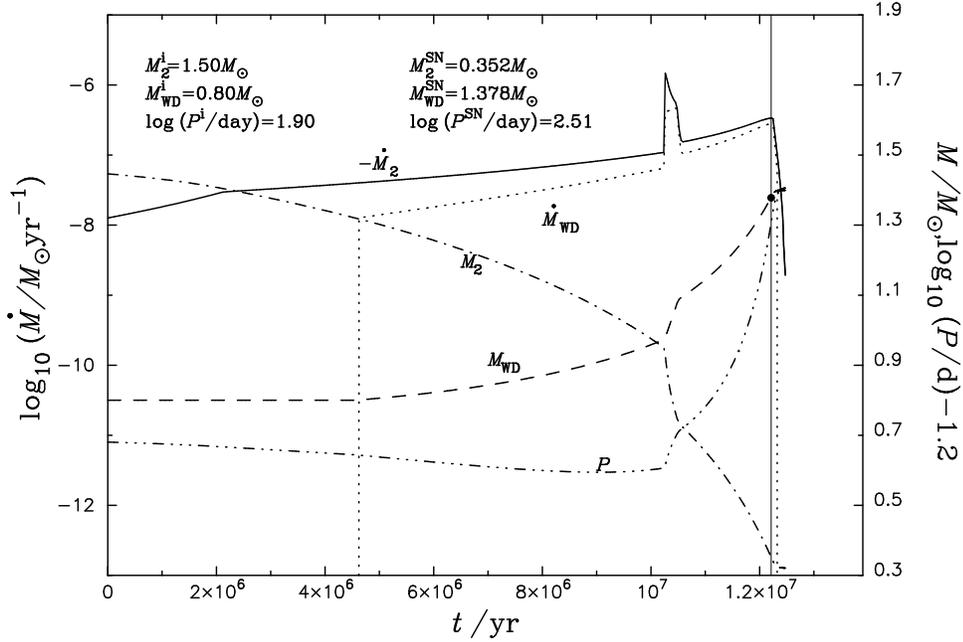}
\caption{The evolution of a binary system that leads to a SN~Ia at the
point marked with the filled circle on the vertical line just after
$1.2\times 10^7\,$yr, under
the companion reinforced attrition process with
$B_{\rm w}=10{,}000$.
The time is arbitrarily offset by $2.666 \times
10^9$\,yr from the birth of the system so that the timescale for the
accretion process is more easily seen in this figure.
The evolution of the secondary
mass $M_{\rm 2}$, the CO WD mass $M_{\rm WD}$, the mass loss
rate from the secondary $-\dot{M}_{\rm 2}$, the mass-growth rate of the
CO~WD $\dot{M}_{\rm WD}$ and the orbital period, $\log_{10}(P/{\rm d})-1.2$ are shown.  For this binary, which
is an extreme case (see discussions in the text),
the mass-transfer rate $\dot M_{\rm t}=|\dot{M}_{\rm 2}|$.  The peak at $t
\approx 1.02\times10^7 {\rm yr}$ marks a phase of stable RLOF.  The CO~WD
growth in other models is similar but most have wind accretion only
before RLOF. \label{example}}
\end{figure}

In Figure~\ref{example} we show the evolution of a binary system
which ends as a SN~Ia.  Initially $M_{\rm WD}^{\rm i}=
0.8\,M_\odot$, $M_{\rm 2}^{\rm i}= 1.5\,M_\odot$ and
$\log_{10}(P^{\rm i}/{\rm d}) = 1.9$.  The wind enhancement factor
$B_{\rm W}=10,000$.  We see three phases of CO~WD growth to
$1.378\,M_\odot$ at which carbon ignites degenerately.  First,
wind accretion becomes important at about $4.7\times 10^6\,$yr
when the accretion is sufficient for hydrogen to burn as it
accretes.  Stable mass transfer dominates for a short time of
around $1.02\times 10^7\,$yr.  Finally wind accretion continues
after the system has detached because the orbit grows faster than
the giant.  The supernova occurs when the secondary's mass has
fallen to $0.353\,M_\odot$ just as it is about to leave the red
giant branch and shrink to a white dwarf.

At $2.673 \times 10^9$\,yr ($t \approx 7
\times 10^6$\,yr in Figure~\ref{example}) the enhanced mass-loss rate
reaches its maximum of 150~times Reimers' rate.  In this case the WD
is able to accrete all the wind and $\dot {M}_{\rm 2a}=-\dot M_{\rm
2w}$.  This is an extreme case of small $P^{\rm i}$, corresponding to
the short orbital period edge of the space that leads to SNe~Ia for an
initially $0.8\,M_\odot$ CO~WD (see Figure~\ref{grid} below).  At
longer initial periods the WD reaches $1.378\,M_\odot$ before RLOF
occurs so there is no stable mass transfer phase.  For much wider
systems and for more massive secondaries, that ignite helium
non-degenerately and consequently do not grow very large on their
first ascent of the giant branch, similar evolution occurs when the
secondary is on the asymptotic giant branch.

Figure~\ref{grid} shows the results of our binary evolution
calculations in an initial period--secondary mass plane for various
initial WD masses.  No SNe~Ia arise when $M_{\rm WD}^{\rm
  i}=0.6\,M_\odot$.  Whether a CO WD reaches the critical mass to
explode as a SN~Ia depends on both the amount of mass transferred and
the efficiency at which this can be accreted by the CO WD.  In each
panel contours enclose the progenitors of SNe~Ia.  Only for the very
high initial white dwarf masses, of $1.1\,M_\odot$, can a massive
secondary star drive a SN~Ia.  These stars ignite helium
non-degenerately and so do not grow large enough to transfer
sufficient mass while on their first ascent of the giant branch.
Thus, only those initially massive enough to retain enough envelope
mass to transfer via a wind on the first ascent of the giant branch
are able to reach SNe~Ia.  In this case the initial mass of the
secondary ought not to be larger than the progenitor of the WD.  This
determines our upper limit for the upper region in panel~(d) and
explains why this region is absent for lower-mass WDs.  The other
regions are more interesting because they are more populated by virtue
of the stellar mass function.  For these the low secondary mass
boundary is due to the limited amount of mass available for transfer
from the lower-mass companions. The long period boundary is due to the
efficiency of burning and accumulating the transferred material.  The
high secondary mass and low period boundary are both due to the onset
of RLOF and dynamically unstable mass transfer because CRAP has not
removed enough mass.  Carbon burning then begins more gently in the
heated outer layers of the WD and a supernova is avoided.

\begin{figure}
\includegraphics[width=6.5cm,angle=270]{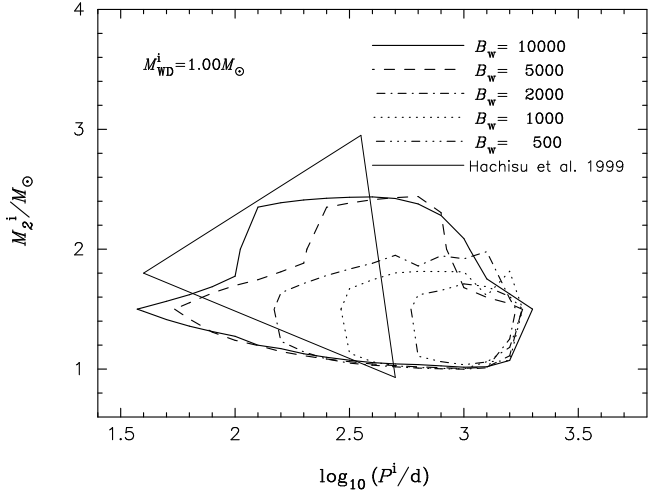}
\caption{Contours for SNe~Ia in the (${\rm log} P^{\rm
i}$, $M_{\rm 2}^{\rm i}$) plane for a $1\,M_\odot$ CO WD with various
values of $B_{\rm w}$.  The thin solid triangle roughly encloses the
SNe~Ia progenitors in the mass-stripping scenario of \citet{hac99a}.
\label{1p00}}
\end{figure}

To examine the effect of CRAP in more detail we repeated our
calculations for an initially $1\,M_\odot$ WD with $0 < B_{\rm w}
< 10{,}000$ (see Figure~\ref{1p00}).  As $B_{\rm w}$ decreases the
progenitor region shrinks and vanishes as $B_{\rm w}\rightarrow
0$. This is consistent with the claim of \citet{iben1984} that
there is no symbiotic channel to SNe~Ia.  For comparison, the
progenitor region found by \citet{hac99a} is also shown in this
figure.  Our model extends the possible progenitor periods by
almost a factor of $\sim 4-5$.

The velocity $v_{\rm w}$ of the companion's wind is
important because it strongly affects $\dot M_{\rm 2a}$.  However
the growth rate of the CO WD $\dot M_{\rm WD}$ is not always
sensitive to $v_{\rm w}$ because $\dot M_{\rm WD}$ is limited by
$\dot M_{\rm cr}$ when the mass accretion rate is large enough
(equation~\ref{mdotcrit1}).  Thus once $\dot M_{\rm 2w}$ is larger
than about $10^{-6}\,M_\odot\,{\rm yr}^{-1}$ the dependence of $\dot
M_{\rm WD}$ on $v_{\rm w}$ becomes small because only about
10\,percent or less of $\dot M_{\rm 2w}$ is required for the CO WD
mass to increase at $\dot M_{\rm cr}$.  When $\dot M_{\rm 2w}$ is
closer to $10^{-7}\,M_\odot\,{\rm yr}^{-1}$ the CO WD may not
increase at all if $v_{\rm w}$ is large.  Our choice of $v_{\rm
w}=5\,{\rm km\,s^{-1}}$ is an estimate of the lower limit to the poorly
modelled wind velocity.  We choose it to demonstrate
that such a channel to SNe~Ia can exist but note that our
progenitor region would shrink in places if $v_{\rm w}$
were larger\footnote{We investigated two alternatives in a little
more detail (a) $v_{\rm w}=10\,{\rm km\,s^{-1}}$ and (b) $v_{\rm
w}=\alpha_{\rm w}v_{\infty}$, where $v_{\infty}$ is a terminal
wind velocity of typically $15\,{\rm km\,s^{-1}}$ for a giant star
and $\alpha_{\rm w}$ is calculated according to \citet{yun95}.  In
both cases the progenitor regions shrink with the longest period
systems not reaching sufficient WD masses to ignite.  For example
the upper period boundary moves to $\log_{10}(P/{\rm d}) \ge 2.9$ for
$M_{\rm WD}^{\rm i} = 1.0\,M_\odot$ and no SNe~Ia are predicted at
all when $M_{\rm WD}^{\rm i} = 0.65\,M_\odot$ in case~(a).}.

Figure~\ref{m2p-sn} shows the regions in the final period--secondary
mass space at which the WD reaches $1.378\,M_\odot$ for various
initial CO WD masses.  The region for a $1\,M_\odot$ CO~WD calculated
with the mass-stripping effect \citep{hac08} is superimposed in the
figure.  As for the initial systems, these regions extend to longer
orbital periods.  Two symbiotic stars, that are considered possible
progenitors of SNe~Ia, T~CrB and RS~Oph, with orbital periods 228\,d
\citep{belc98} and 454\,d \citep{brandi09}, are plotted under the
assumption that $M_{\rm WD} \approx 1.378\,M_\odot$.  Both are located
well inside the region of SNe~Ia progenitor regions for our models but
lie on the edge of what might be expected with mass stripping.

\section{Frequency estimates}

\begin{figure*}
\includegraphics[width=12cm,angle=270]{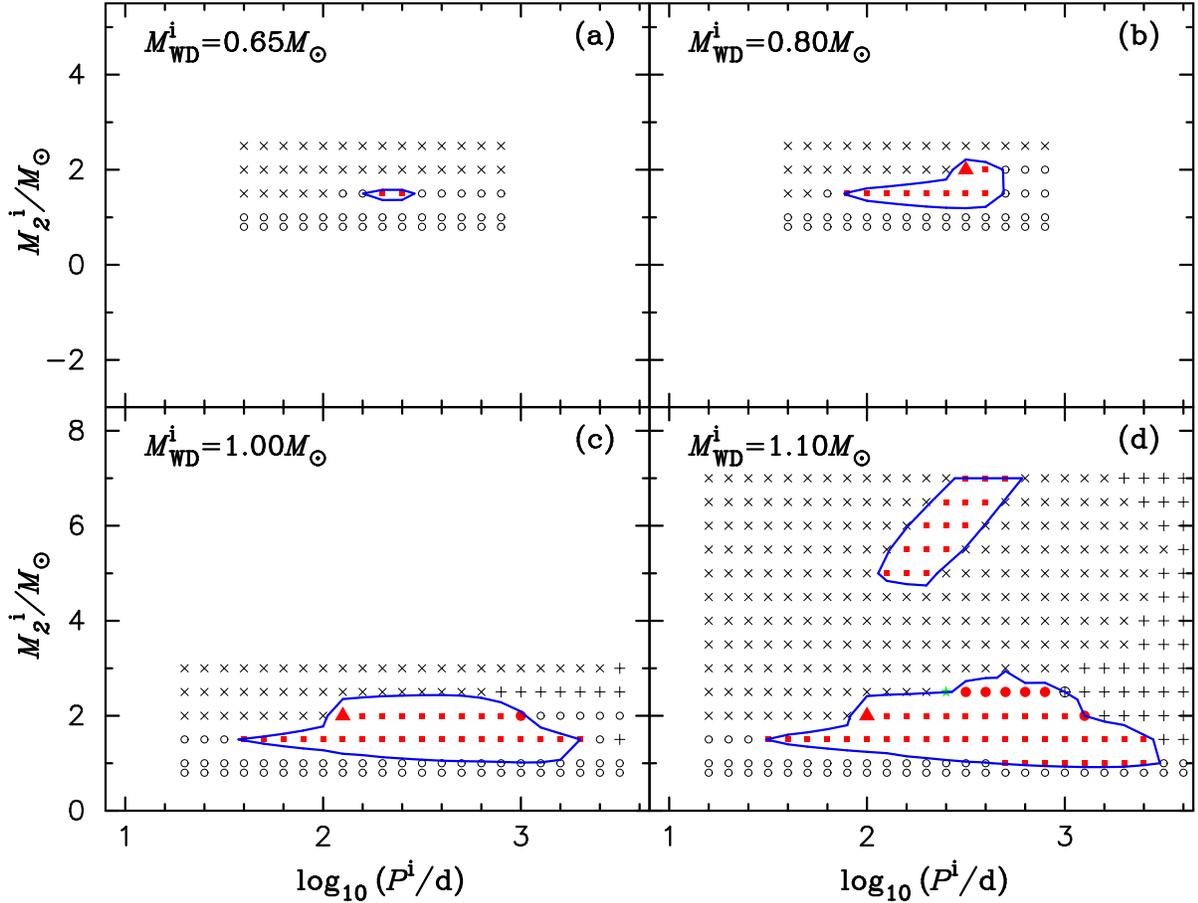}
\caption{Results of binary evolution calculations in (${\rm log}
P^{\rm i}$, $M_{\rm 2}^{\rm i}$) planes, where $B_{\rm w}=10000$
for four initial masses $M_{\rm WD}^{\rm i}$.  Initial periods are
$P^{\rm i}$ and initial secondary masses are $M_{\rm 2}^{\rm i}$.
Filled symbols indicate the progenitors of SNe~Ia explosions.  For
filled triangles and squares the explosion occurs before helium
ignition in the secondary star. For the triangles the explosion
occurs during RLOF while for the squares the explosion occurs in
the wind accretion phase. The final companion remnant would be a
helium white dwarf in these cases.  Filled circles indicate
systems that explode as SNe~Ia after central helium burning and
before RLOF. In these cases the final companion remnants would be
CO WDs.  Symbols `$\times$' and stars indicate systems that are
unstable to dynamical mass transfer with CO WD masses lower or
higher than $1.35M_\odot$ respectively but still less than
$1.378\,M_\odot$. Open circles are systems stable to dynamical
mass transfer but in which there is insufficient mass for the CO
WDs to reach $1.378M_\odot$. Systems marked `+' and `$\rm
\bigoplus$' begin AGB thermal pulses before RLOF or a SN~Ia. It
becomes computationally difficult and inefficient to evolve these
stars fully so we make a simple estimate of whether a SN~Ia is
likely. Only in the system marked `$\rm \bigoplus$' is the CO WD
likely to reach $1.378M_\odot$. The heavy contours enclose the
systems in which the WD can explode as a SN~Ia. \label{grid}}
\end{figure*}

\begin{figure}
\includegraphics[width=6.5cm,angle=270]{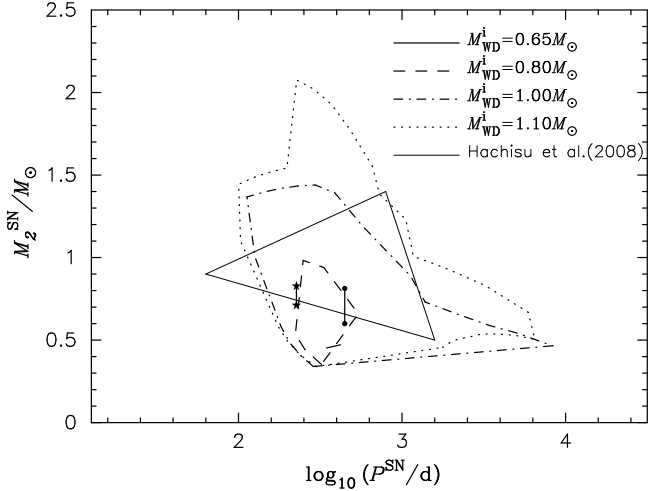}
\caption{Contours enclosing progenitors of SNe~Ia, now in the final
period--secondary mass plane just before explosion for various initial
WD masses and also the triangle for \citet{hac99a}'s mass-stripping
scenario.  Filled stars and circles are the current positions of two
symbiotic binary stars, T CrB and RS Oph.  To plot these we have
assumed both the CO~WDs have a mass of $1.378\,M_\odot$ so that the
secondary mass for T~CrB is $0.71\,M_\odot$ according to
\citet{hac99a} or $0.823\,M_\odot$ derived from the mass ratio of
about $0.6$ determined by \citep{belc98} and for RS~Oph
$0.68-0.8M_\odot$ according to
\citet{brandi09}.  As in panel~(d) of Figure~\ref{grid}
the contour for $1.1\,M_\odot$ has two parts but the
upper region is off this figure.
\label{m2p-sn}}
\end{figure}

Figure~\ref{grid} shows how the CRAP enhanced symbiotic channel
would be very effective if many binary stars with a massive CO~WD and
an unevolved companion could form with periods between about $100$~and
$1{,}000\,$d.  However conventional population synthesis calculations
tend to produce such systems either with $P<100\,$d following the
ejection of a common envelope or with $P>1{,}000\,$d if they have
avoided RLOF.  On the other hand we do observe symbiotic stars, such
as T~CrB and RS~Oph, with massive WDs and orbital periods around 1\,yr.
\citet{hac99a} describe one way in which these systems, and
consequently many like them, could have formed by shrinking a wide
orbit in a process similar to common envelope evolution but by
interaction with the stellar wind.
To obtain the frequency of SNe~Ia
produced by our channel we really ought to carry out a full population
synthesis from the zero-age main sequence but here we apply the
procedure explained by \citet{hac99a} in their sections~4.3
and~4.4 so as to get a direct comparison with their rate.  We
divide the WD masses into the ranges given in our
Table~\ref{table} and calculate the integral in equation~(1) of
\citet{iben1984},
\begin{equation}
\nu = 0.2\,{\rm yr^{-1}}\Delta q\Delta\log_{10}a \int_{M_{\rm
A}}^{M_{\rm B}}M^{-2.5}\,dM,
\label{tutukov}
\end{equation}
where $M_{\rm A}$ and $M_{\rm B}$ are the typical progenitor masses of
the WDs at each end of the range.  The integral then accounts for
the initial mass function.  We calculate the initial range of mass
ratios, $\Delta q$, for a given WD mass by
\begin{equation}
\Delta q = \frac{M_{\rm u}}{M_{\rm A}} - \frac{M_{\rm l}}{M_{\rm B}},
\end{equation}
where $M_{\rm u}$ and $M_{\rm l}$ are the upper and lower limits
to the SNe~Ia progenitor regions identified directly from
Figure~\ref{grid}.  By formula~(\ref{tutukov}) we assume that the
mass ratio is uniformly distributed between $0$ and~$1$.  With the
quantity $\Delta\log_{10}a$ we assume that the initial period
distribution is flat in $\log P$ and that this difference is
unchanged by the evolution up to the point of Figure~\ref{grid} so
that we may read it from the figure directly for the particular WD
mass.  We then calculate a simple mean rate $\bar\nu$ of the two
extremes.  Our total SNe~Ia rate is $0.0069\,{\rm yr^{-1}}$,
significantly larger than that calculated by \citet{hac99a} with
their mass-stripping.  This rate is very uncertain and probably
overestimated because we, as did they, have used cuboid shaped
regions in the initial parameter space but it is not dissimilar to
the actual observed rate for a galaxy like our own.

\begin{table}
 \begin{minipage}{120mm}
 \caption{
\label{table} The frequency of SNe~Ia from our model with $B_{\rm
w}=10{,}000$.  For each range of WD masses $M_{\rm WD}$, $M_{\rm
A}$ and $M_{\rm B}$ are the corresponding lower and upper zero-age
progenitor masses and $\Delta q$ and $\Delta\log_{10}a$ are the
typical mass ratio and separation ranges for the regions in
Figure~\ref{grid} which lead to SNe~Ia.  The frequencies $\nu$ and
$\bar\nu$ are the calculated supernova rates if the
extremes of the distribution apply across the whole range and a
simple mean of these two extremes.  The factors of $2/3$ account
for the change from period to separation space.
\label{tab1}}
   \begin{tabular}{ccccccc}
\hline
$M_{\rm WD/M_\odot}$&$\Delta\log_{10}a$&$M_{\rm
A}/M_\odot$&$M_{\rm B}/M_\odot$&$\Delta q$&$\nu/{\rm yr^{-1}}$&$\bar{\nu}/{\rm yr^{-1}}$\\
\hline
0.65-0.8 &$0.3\times \frac{2}{3}$&2.42&4.48&0.3168&0.0013&0.0043\\
        &$0.8\times \frac{2}{3}$&2.42&4.48&0.6412&0.0073&\\
\hline
0.8-1.0 &$0.8\times \frac{2}{3}$&4.48&6.63&0.3101&0.0001&0.0019\\
        &$1.7\times \frac{2}{3}$&4.48&6.63&0.3960&0.0028&\\
\hline
1.0-1.1 &$1.7\times \frac{2}{3}$&6.63&7.58&0.2376&0.0004&0.00055\\
(lower contour)&$2.0\times \frac{2}{3}$&6.63&7.58&0.3787&0.0007&\\
\hline
1.0-1.1 &0&-&-&-&0.0000&0.00015\\
(upper contour)&$0.7\times \frac{2}{3}$&6.63&7.58&0.4489&0.0003&\\
\hline
\label{frequency}
\end{tabular}
\end{minipage}
\end{table}

\acknowledgments We are grateful to the anonymous referee for
his or her valuable comments.  This work is
supported by the NSFC (Nos. 10973036, 11033008, 10821061 and
2007CB815406), the CAS (No. KJCX2-YW-T24) and Yunnan National
Science Foundation (No. O8YJ041001).  XC also thanks the
Talent Project of Western Light supported by the CAS.  CAT thanks
Churchill College for his Fellowship.

\clearpage

\begin{thebibliography}{}
\bibitem[Anupama \& Miko{\l}ajewska(1999)]{anu99}
 Anupama G. C., \& Miko{\l}ajewska, J. 1999, A\&A, 344, 177
\bibitem[Belczy\'nski \& Miko{\l}ajewska(1998)]{belc98}
 Belczy\'nski, K., \& Miko{\l}ajewska, J. 1998, MNRAS, 296, 77
\bibitem[Boffin \& Jorissen(1988)]{boff88}
 Boffin, H. M. J., \& Jorissen A. 1988, A\&A, 205, 155
\bibitem[Bona\v{c}i\'c, Glebbeek \& Pols(2008)]{bon08}
 Bona\v{c}i\'c M., Glebbeek E., \& Pols, O. R. 2008, A\&A, 480, 797
\bibitem[Botticella et al.(2008)]{bot08}
  Botticella, M. T., et al. 2008, A\&A, 479, 49
\bibitem[Brandi et al.(2009)]{brandi09}
  Brandi, E., Garc\'\i a, L. G., Quiroga, C., Ferrer, O. E., \& Marchiano, P. 2009, A\&A, 497, 81
\bibitem[\protect\citeauthoryear{Davidson \&
Ostriker}{1973}]{davidson1973}Davidson, J., \& Ostriker, J. P. 1973,
179, 585
\bibitem[Eggleton(1971)]{egg71}
  Eggleton, P. P. 1971, MNRAS, 151, 351
\bibitem[Gilfanov \& Bogd\'an(2010)]{gilf10}
Gilfanov, M., \& Bogd\'an, A. 2010, Nature, 463, 924
\bibitem[Hachisu, Kato \& Nomoto(1999a)]{hac99a}
  Hachisu, I., Kato, M., \& Nomoto, K. 1999a, ApJ, 522, 487
\bibitem[Hachisu et al.(1999b)]{hac99b}
  Hachisu, I., Kato, M., Nomoto, K., \& Umeda H. 1999b, ApJ, 519, 314
\bibitem[Hachisu, Kato \& Nomoto(2008)]{hac08}
  Hachisu, I., Kato, M., \& Nomoto, K. 2008, ApJ, 683, L127
\bibitem[Hachisu, Kato \& Nomoto(2010)]{hac10}
  Hachisu, I., Kato, M., Nomoto, K. 2010, ApJ, 724, L212
\bibitem[Han(1998)]{han98}
  Han, Z. 1998, MNRAS, 296, 1019
\bibitem[Han \& Podsiadlowski(2004)]{han04}
  Han, Z., \& Podsiadlowski, P. 2004, MNRAS, 350, 1301
\bibitem[\protect\citeauthoryear{Hurley, Tout \&
Pols}{2002}]{hurley2002}Hurley, J. R., Tout, C. A., \& Pols,
O. R. 2002, MNRAS, 329, 897
\bibitem[Iben \& Tutukov(1984)]{iben1984}
  Iben, I. Jr., \& Tutukov, A. V. 1984, ApJs, 54 , 335
\bibitem[\protect\citeauthoryear{Kenyon \&
Webbink}{1984}]{kenyon1984}Kenyon, S. J. \& Webbink, R. F. 1984, ApJ,
279, 252
 \bibitem[Meng, Chen \& Han(2009)]{meng09}
  Meng, X., Chen, X., \& Han Z., 2009, MNRAS, 395, 2103
\bibitem[Perlmutter et al.(1999)]{per99}
 Perlmutter, S., et al. 1999, ApJ, 517, 565
\bibitem[Pols et al.(1998)]{pol98}
 Pols, O. R., Schr\"oder K.-P., Hurley J. R., Tout C. A., \& Eggleton
P. P. 1998, MNRAS, 298, 525
\bibitem[\protect\citeauthoryear{Reimers}{1975}]{reimers1975}Reimers,
D. 1975, M\'em. Roy. Soc. Li\`ege, 8, 369
\bibitem[Riess et al.(1998)]{riess98}
Riess, A., et al. 1998, AJ, 116, 1009
\bibitem[Schawinski(2009)]{sch09}
Schawinski, K. 2009, MNRAS, 397, 717
\bibitem[Totani et al(2008)]{tot08}
Totani, T., Morokuma, T., Oda, T., Doi, M., \& Yasuda N. 2008, PASJ, 60,
1327
\bibitem[Tout \& Eggleton(1988a)]{tout88a}
Tout, C. A., \& Eggleton, P. P. 1988a, MNRAS, 231, 823
\bibitem[Tout \& Eggleton(1988b)]{tout88b}
Tout, C. A., \& Eggleton, P. P., 1988b, ApJ, 334, 357
\bibitem[van den Heuvel et al.(1992)]{heu92}
van den Heuvel, E. P. J., Bhattacharya, D., Nomoto, K., Rappaport,
S. 1992, A\&A, 262, 97
\bibitem[van Winckel(2003)]{winck03}
van Winckel, H. 2003, ARA\&A, 41, 391
\bibitem[Wang et al.(2009)]{wang09}
Wang, B., Meng, X., Chen, X., \& Han Z., 2009, MNRAS, 395, 847
\bibitem[Yungelson et al(1995)]{yun95}
Yungelson, L., Livio, M., Tutukov, A., \& Kenyon, S. J. 1995, ApJ, 447, 656
\end{thebibliography}
\end{document}